\documentclass{pnastwo}

\usepackage{graphicx}
\usepackage{amssymb,amsfonts,amsmath}
\usepackage{color} 

\newcommand{\ep}{\epsilon}
\newcommand{\bep}{{\bf \epsilon}}
\newcommand{\bp}{{\bf p}}
\newcommand{\glog}{\Lambda}
\newcommand{\rr}{x_0}
\newcommand{\gf}{h}

\begin{document}


\title{Generalized entropies and the transformation group of superstatistics}

\author{Rudolf Hanel\affil{1}{Section for Science of Complex Systems, Medical University of Vienna, Spitalgasse 23, 1090 Vienna, Austria}, 
Stefan Thurner
\affil{1}{}\affil{2}{Santa Fe Institute, 1399 Hyde Park Road, Santa Fe, NM 87501, USA}
\and Murray Gell-Mann\affil{2}{}}


\maketitle


\begin{article}

\begin{abstract} 
Superstatistics describes statistical systems that behave like superpositions of different inverse temperatures $\beta$, so that the probability distribution is 
$p(\ep_i) \propto \int_{0}^{\infty} f(\beta) e^{-\beta \ep_i }d\beta$, where the `kernel' $f(\beta)$ is 
nonnegative and normalized ($\int f(\beta)d \beta =1$).   We discuss the relation between this distribution and the generalized entropic form $S=\sum_i s(p_i)$. The first three Shannon-Khinchin axioms are assumed to hold. It then turns out that for a given distribution there are two different 
ways to construct the entropy. One approach uses escort probabilities and the other does not; the question of which to  use must be decided empirically. 
The two approaches are related by a duality.
The thermodynamic properties of the system can be quite different for the two approaches. In that connection we present the transformation laws for the superstatistical distributions under macroscopic state changes. The transformation group is the Euclidean group in one dimension. 
\end{abstract}

\keywords{thermodynamics | entropy | classical statistical mechanics | correlated systems}

\dropcap{S}uperstatistics  \cite{beck03} has been introduced as a way of systematically handling statistical systems 
that can be seen as superpositions of various Boltzmann distributions. One may think of a superstatistical system on 
a microscopic and a macroscopic scale. At the microscopic scale the system relaxes towards thermodynamic equilibrium 
and follows a single Boltzmann statistic with a well defined {\em local} inverse temperature $\beta$, i.e. the local 
probability of finding the system at  some  energy $\ep$ is proportional to $\exp(-\beta\ep)$. 
On the macroscopic scale the local inverse temperatures fluctuate, with the fluctuations governed by  $f(\beta)$
\footnote{Superstatistical systems are usually not in equilibrium but in stationary non-equilibrium. 
		E.g., think of a gas or liquid between two plates at different temperatures $T_1$ and $T_2$. 
		\label{foo_T12gas}}.
Following \cite{beck03}, the generalized superstatistical Boltzmann factor reads
\begin{equation}
	B(\ep)=\int_0^\infty d\beta\, f(\beta)e^{-\beta \ep}  \quad ,
\label{superstat}
\end{equation} 
where $f(\beta)$ is called the {\it superstatistical kernel}. The kernel is non-negative and normalized, i.e. $\int_0^\infty d\beta \, f(\beta)=1$.
The probability of finding the system in one of $W$ discrete (for example energy) states, 
$\bep=\{\ep_i\}_{i=1}^W$, is 
\begin{equation}
	p_i= \frac{1}{Z}B(\ep_i)\quad,\quad Z=\sum_{i=1}^W B(\ep_i)\,,
\label{superdistr}
\end{equation} 
$Z$ being the superstatistical generalization of the partition function. The expectation value 
\begin{equation}
	U=\sum_{i=1}^W p_i \ep_i\,
\label{expectedenergy}
\end{equation} 
is the internal energy of the system.

Superstatistics has found many applications ranging from hydrodynamic turbulence \cite{beck05,beckEPL03,reynolds03},  
complex networks  \cite{abeNW}, and  pattern formation  \cite{daniels04} to finance \cite{bouchaud,hase03}.
Besides these practical aspects, superstatistics has the potential for providing a structural foundation for non-Boltzmann 
statistical mechanics. 

Different methods have been proposed for establishing the mathematical methodology of a kind of statistical mechanics compatible with superstatistics \cite{TsallisSouza,abe_ss_03,abc07,hanel07, thurner08}. 
The methods that use generalized entropic forms are the Tsallis Souza (TS) approach \cite{TsallisSouza} and 
one described in \cite{hanel07, thurner08} which we refer to as the HT approach. 
Both  use a maximum entropy principle (MEP) to {\em reconstruct} generalized 
entropies having the structure
\begin{equation}
	S[\bp]=\sum_{i=1}^W s(p_i) 
\label{ents}
\end{equation}
from given superstatistical distribution functions. Here $\bp$ stands for the set of probabilities $\{p_i\}_{i=1}^W$.
The sum structure is chosen in analogy to Boltzmann-Gibbs (BG) entropy, where $s_{BG}(x)=-x\log(x)$.
The function $s$ is continuous and concave and has $s(0)=0$. These properties correspond
to the first three Shannon-Khinchin axioms\footnote{Shannon-Khinchin axioms:
	(i) Entropy is a continuous function of the probabilities $p_i$ only, i.e. $s$ should not explicitly depend on 
	any other parameters. 
	(ii) Entropy is maximal for the equi-distribution $p_i=1/W$.
	-- From this the concavity of $s$ follows. 
	(iii) Adding to a system a state numbered $W+1$ with $p_{W+1}=0$ does not change the entropy of the system. 
	-- From this $s(0)=0$ follows.
	\label{foo_shannon}}
 \cite{shannon,Kinchin}. 
In this paper we address the question of whether the TS approach is the unique way to reconstruct entropies for superstatistical systems.  
The answer is no -- there exist possible alternatives. In fact, if we did not have the first Khinchin axiom we would 
have infinitely many possibilities.  We show however, that assuming this axiom we have only two possibilities remaining. 
The axiom requires that the entropy be a continuous function of the probabilities $\bp$ only 
(compare footnote \ref{foo_shannon}). It turns out that one possibility corresponds to the TS  \cite{TsallisSouza} and the 
other to the HT approach  \cite{hanel07, thurner08}.  We show that the HT case cannot be 
seen as a simple limit 
of the TS case. 
For a large class of distribution functions 
such a limit yields the BG case, not HT in general. 
The thermodynamic properties of the entropies 
corresponding to the two approaches will in general be quite different. Which approach is the 
correct one to use in each case has to be decided empirically. 
Superstatistics -- as originally presented--  does not make any assumptions about how the superstatistical 
Boltzmann factor $B$ or the kernel $f$ transform under thermodynamic state changes. 
On the macroscopic scale, a superstatistical system can be described by a macro-state $\sigma$,
in complete analogy to a classical thermodynamic system. 
These states are characterized by a set of macro-variables, e.g.  internal energy, volume, temperatures,  $\sigma=\{U,V,T,...\}$. 
We present the transformation group of the superstatistical kernel $f$ under changes from one state to another, $\sigma \to \sigma'$.

\section{Entropy reconstruction from distribution functions}

The usual procedure when using the MEP is to maximize a given entropy under given constraints, represented 
by Lagrange multipliers:
\begin{equation}
	\Phi[\bp]=S[\bp]-\alpha \left(\sum_i p_i-1\right)-\beta\left(\sum_i p_i\ep_i-U\right) \quad .
\label{HTfunct}
\end{equation} 
Given $s(p_i)$ and $U$,  this yields the distribution function $p_i$, along with the values of the Lagrange multipliers 
$\alpha$ and $\beta$.    

In the case of entropy reconstruction the distribution function of a superstatistical system is given. The system is in a specific 
physical state\footnote{The state $\sigma$ does not change during the reconstruction.} $\sigma$. The various quantities such 
as $p_i^{(\sigma)}$, the internal energy $U^{(\sigma)}$, etc. are given and are labeled by $(\sigma)$. Since $p^{(\sigma)}$ is a 
superstatistical distribution, we have $p_i^{(\sigma)}= B(\ep^{(\sigma)})/Z^{(\sigma)}$. The quantity  $s(p_i)$ and  the entropy can now be computed 
for the state $\sigma$.   

\subsection{Entropy reconstruction --  the HT approach}

We begin with the simpler case and demonstrate the entropy reconstruction using the HT approach \cite{hanel07,thurner08}, 
which starts by maximizing the functional of  Eq. (\ref{HTfunct}).  At this point we use the available information about the specific state $\sigma$. 
The condition 
\begin{equation}
	\partial \Phi[{\bf p} ^{(\sigma)}] /\partial p^{(\sigma)}_i=0 
\end{equation} 
leads to
\begin{equation}
	s_{HT}'(p^{(\sigma)}_i)=\alpha^{(\sigma)} + \beta^{(\sigma)} \ep^{(\sigma)}_i \quad.
\label{HTDGLsig}
\end{equation}
Since $B^{(\sigma)}(\ep)$ is a monotonic function of $\ep$ the inverse function $L$ defined by  
\begin{equation}
	L\left(\frac{B^{(\sigma)}(\ep)}{Z^{(\sigma)}}\right)=\ep 
\label{inverseHT}
\end{equation}
exists. Note that for the specific state $\sigma$,  $Z^{(\sigma)}$ is just a positive real number. Otherwise the inverse 
would be hard to calculate. It follows immediately that 
\begin{equation}
	L\left(p^{(\sigma)}_j\right)=\ep^{(\sigma)}_j  
\label{inverse2HT}
\end{equation}
for all $j$.
Equations  (\ref{HTDGLsig}) and (\ref{inverse2HT}) yield the differential equation
\begin{equation}
	s_{HT}'\left(p^{(\sigma)}_i\right)=\alpha^{(\sigma)} + \beta^{(\sigma)} L\left(p^{(\sigma)}_i\right) \quad. 
\label{HTDGL_2}
\end{equation}
This allows us to reconstruct the entropic form by integration:  
\begin{equation}
	s_{HT}(x)= \alpha^{(\sigma)} x + \beta^{(\sigma)} \int_{0}^{x} dy L(y) \quad. 
\label{HT}
\end{equation}
We have now derived an entropic form\footnote{
	In \cite{hanel07, thurner08} $\Lambda(x)=- \alpha -\beta L(x)$ is called the generalized logarithm. 
} 
$s_{HT}$ which, when used in the MEP, Eq. (\ref{HTfunct}), reproduces
$p^{(\sigma)}_i$, $\alpha^{(\sigma)}$, and $\beta^{(\sigma)}$ as a solution, given the state characterized by 
$U^{(\sigma)}$ and $\ep^{(\sigma)}_i$.  As a result of the first Khinchin axiom this entropic form $s_{HT}$ characterizes 
the superstatistical system in general, i.e. not only for the state $\sigma$ but also for the other possible states $\sigma'$,  
corresponding  to different values of $U^{(\sigma')}$ or $\ep_i^{(\sigma')}$. This has consequences for the thermodynamical 
treatment of the system. In particular, transformation laws for the superstatistical Boltzmann factor $B$ and the kernel $f$ 
under changes of the state $\sigma$ can be derived, as will be shown below.

\subsection{Entropy reconstruction with escort distributions -- the TS approach}

We now generalize the above idea by modifying the energy constraint.  We compute a different  `energy'  
$U^*$ by using a different set of  `probabilities'. These probabilities can be written in the general  form 
\begin{equation}
	 P_i= \frac{u(p_i )}{\sum_j u(p_j)} \quad.
	 \label{constraint}
\end{equation}   
The quantity $P_i$ is often called {\em escort probability}. The escort energy $U^*$ is identical with the internal energy $U$ if and only if  
$u(p_i)=p_i$. We do not attempt to provide a physical interpretation of the escort probabilities. We start with the functional 
\begin{equation}
	\Phi[\bp]=S[\bp]-\alpha \left(\sum_i p_i-1\right)-\beta\left(\sum_i P_i\ep_i-U^*\right) \quad . 
\label{TSfunct}
\end{equation} 
For  taking the derivative of $P$, we define the following notation using a function $Q$ of two variables: 
\begin{equation}
\begin{array}{ccl}
	u(p_i) &\equiv& Q[x,y] \mid_{x=p_i; y=s(p_i)}   \\
 	u_{1}(p_i) &\equiv&   \frac{\partial}{\partial x} Q[x,y]\mid_{x=p_i; y=s(p_i)} \\
	 u_{2}(p_i) &\equiv&  \frac{\partial}{\partial y} Q[x,y] \mid_{x=p_i; y=s(p_i)} \quad .
\end{array}
\end{equation}
Maximizing the functional in Eq. (\ref{TSfunct}) we have $\partial \Phi/\partial p_i=0$, leading to
\begin{equation}
	s'(p_i)=\alpha + \beta^* (\ep_i-U^*)\left[u_1(p_i)+u_2(p_i)s'(p_i)\right]\quad,
\label{TSDGL_0}
\end{equation}
where $\beta^*$ is the rescaled parameter 
$\beta/\sum_i u(p_i)$. 
Solving for $s'$, we get 
\begin{equation}
	s'(p_i)=\frac{\alpha + \beta^* (\ep_i-U^*)u_1(p_i)}{1-\beta^*(\ep_i-U^*)u_2(p_i)}\quad.
\label{TSDGL}
\end{equation}
As in the HT approach, we  consider from this point on a specific state $\sigma$ of the superstatistical system and use 
the specific values $\alpha^{(\sigma)}$, $\beta^{(\sigma)}$, $\dots$, for reconstructing the associated
entropy  $s$. However, for the sake of readability we suppress $(\sigma)$ whenever that cannot cause confusion. 
Using Eq. (\ref{inverse2HT}) as above,  we insert $\ep_i=L(p_i)$ into 
Eq. (\ref{TSDGL})  and get 
\begin{equation}
	s'(x)=\frac{\alpha + \beta^* (L(x)-U^*)u_1(x)}{1-\beta^*(L(x)-U^*)u_2(x)}\quad.
\label{TSDGL_2}
\end{equation}
This differential equation in $s$ can now be solved, so that for any suitable function $u$ an entropic form $s$ is 
obtained which reproduces the superstatistical distribution function $p^{(\sigma)}_i=B^{(\sigma)}(\ep_i)/Z^{(\sigma)}$ under 
the MEP. 

It is easy to see that such entropies could explicitly depend on $U^*$, except for the first  Khinchin axiom, which rules out such dependence. 
This reduces dramatically the possibilities for choosing $u$. 
In fact, the first axiom can hold only if both $u_1(x)=r_1$ and $u_2(x)=r_2$ are constants, implying $u(p_i)=r_1 p_i+r_2 s(p_i)$.  
This now allows us to absorb $U^*$ into the Lagrange multipliers. By defining new constants 
\begin{equation}
	\hat \alpha =\frac{\alpha-\beta^*U^*r_1}{1+\beta^*U^*r_2}\quad {\rm and}\quad
	\hat \beta   =\frac{\beta^*r_1}{1+\beta^*U^*r_2} \quad , 
\end{equation}
we can simplify Eq. (\ref{TSDGL_2}) to 
\begin{equation}
s'(x)=\frac{\hat \alpha + \hat \beta L(x)}{1-\hat \beta\nu L(x)}
\label{TSDGL_3}
\end{equation}
with $\nu=r_2/r_1$. This equation can now be solved by simple integration. Note that since $P_i$ remains invariant if both 
$r_1$ and $r_2$ are rescaled by the same factor  we can use $r_1=1$ without loss of generality. With $u(p_i)= p_i + \nu s(p_i)$, 
the TS approach is recovered. 
Note that more general constraints than those in Eqs. (\ref{HTfunct}) and (\ref{TSfunct}) are ruled out by the first 
Khinchin axiom. 

\section{The two entropies}
We see  that only two entropies remain, one obtained with ordinary constraints (HT approach), and 
the other by employing escort distributions in the functional of  Eq. (\ref{TSfunct}) (TS approach). 

\begin{itemize}

\item {\bf HT entropy:} The approach without escort distributions results in Eq. (\ref{HT})  
\begin{equation}
	s_{HT}(x)=\beta \int_0^x dy L(y)  + \alpha x \quad, 
\label{TSDGL_g}
\end{equation} 
which yields  the HT entropy,  \cite{hanel07, thurner08}.

\item {\bf TS entropy:}  The only possible entropy involving escort distributions results from   Eq. (\ref{TSDGL_3}), which gives 
\begin{equation}
	s_{TS}(x)=\int_0^x dy \frac{\hat \alpha + \hat \beta L(y)}{1-\hat \beta \nu L(y)}\quad,
\label{TSDGL_TS}
\end{equation}
the TS result for the entropy, and $P_i= (p_i+\nu s_{TS}(p_i))/ \sum_j  (p_j+\nu s_{TS}(p_j) )$ for the constraint. 
In \cite{TsallisSouza}  $\hat \beta$ is set equal to one.   
\end{itemize}
Note that the TS approach has the freedom in choosing $\nu$ (each choice of $\nu$ leading to a different entropy $s_{TS}$), whereas in the HT approach the entropy is fully determined by the distribution function alone. 
The question arises of whether there exists a way to unambiguously determine the value of $\nu$. 
That is addressed in the following section. 

\section{Duality of the two approaches \label{superduality}}

Using Eqs. (\ref{TSDGL_g}) and (\ref{TSDGL_TS}) we have for the functional relation between the two entropies  
\begin{equation}
	s_{TS}'(x)=  \frac{\frac{1}{\hat \beta}\left(\hat \alpha \beta - \alpha \hat \beta\right) + s_{HT}'(x) }  
	{\left(\frac{\beta}{\hat \beta} + \nu \alpha\right) -\nu s_{HT}'(x)} \quad. 
\label{duality_main}
\end{equation}
Suppose we reconstruct an entropic form $s$ from state $\sigma$. We now use this form 
in the MEP for another state $\sigma'$.  This could for example be a state with a different  internal energy.
For both approaches, the HT and the TS, solving the MEP for state $\sigma'$ leads to distribution functions of the form
\begin{equation}
	p^{(\sigma')}_i=\frac{1}{Z^{(\sigma)}}B^{(\sigma)}\left( a+b\ep^{(\sigma')}_i \right)\quad ,
\label{dist}
\end{equation}
where for  the HT approach we have 
\begin{equation}
	a=\frac{ \alpha^{(\sigma')}-\alpha^{(\sigma)}  }{  \beta^{(\sigma)}  } \quad {\rm and} \quad 
	b=\frac{\beta^{(\sigma')}}{\beta^{(\sigma)}}\quad ,
	\label{trans_law_ab_HT}
\end{equation}
and for the TS approach 
\begin{equation}
	a=\frac{  \hat \alpha^{(\sigma')} - \hat \alpha^{(\sigma)}   }{ \hat \beta^{(\sigma)} \left(1+\nu \hat \alpha^{(\sigma')} \right) } \quad {\rm and} \quad  
	b=\frac{ \hat \beta^{(\sigma')} }{\hat \beta^{(\sigma)}  }   \frac{  1+\nu\hat \alpha^{(\sigma)}  }{  1+\nu\hat \alpha^{(\sigma')}  }\quad . 
	\label{trans_law_ab_TS}
\end{equation}
From Eqs. (\ref{trans_law_ab_HT}) and  (\ref{trans_law_ab_TS}) and
the requirements $\lim_{\nu\to0}\hat\alpha   =\alpha$ and $\lim_{\nu\to0}\hat\beta=\beta$
we get 
\begin{equation}
0=\hat \alpha \beta - \alpha \hat \beta 
\quad\mbox{and}\quad 
1=\frac{\beta}{\hat \beta} + \nu \alpha \quad,
\end{equation}
so that Eq. (\ref{duality_main}) simplifies to the relation
\begin{equation}
	\frac{1}{s_{TS}'(x)}-\frac{1}{s_{HT}'(x)}=-\nu \quad. 
	\label{duality_1}
\end{equation}
Remarkably $\alpha$ and $\beta$ drop out. 
This relation can be expressed in terms of so-called {\it generalized logarithms} (g-logarithms) 
for the HT and the TS approaches.
Generalized logarithms $\glog(x)$ have been widely used in the context of generalized entropies, e.g. \cite{naudts_physA2002,ht2010}. 
$\glog$ is an increasing monotonic function with $\glog(1)=0$ and  $\glog'(1)=1$.
Equation (\ref{duality_1}) suggests that given $s'_{TS}(x)=0$ for some $x=\rr$, 
$s'_{HT}(\rr)$ is also $0$, and further 
$s''_{TS}(\rr)=s''_{HT}(\rr)$. If we define a constant 
{\bf $c\equiv-\rr\,s_{HT}''(\rr)$}, 
the  
g-logarithms associated with the generalized entropies can be written as 
\begin{equation}
	\glog_{HT}(x)= -\frac1c s_{HT}'(\rr x) \, , \quad \glog_{TS}(x)= -\frac1c s_{TS}'(\rr x) \,. 
	\label{glogs_1}
\end{equation}
With these g-logarithms Eq. (\ref{duality_1})  becomes  the {\em duality relation}
\begin{equation} 
	\glog_{HT}^*(x)=  \glog_{TS} (x) \quad , 
	\label{du}
\end{equation}
where the duality operation $^*$ is defined by 
\begin{equation}
 	\glog^* = \frac{1}{\frac{1}{ \glog}+c\nu}   \quad. 
	\label{duality_2}
\end{equation}
Note that Eq. (\ref{duality_2}) 
possesses a symmetry: the equation is invariant under interchanging $\glog$ ($\glog\equiv\glog_{HT}$) 
with $\glog^*$ ($\glog^*\equiv\glog_{TS}$) and simultaneously changing the sign of $\nu$. Obviously, applying $^*$ a second time (exchanging $\glog_{HT} \leftrightarrow \glog_{TS}$ and $\nu \leftrightarrow -\nu$) yields the identity, as has to be the case for any duality\footnote{
	The duality $^*$ acts on pairs $(\glog,\nu)$ in such a way that 	
	$(\glog,\nu)^*=(\glog^*,\nu^*)$ with $\glog^*$ as in Eq. (\ref{duality_2})
	and $\nu^*=-\nu$. All elements $(\glog,0)$ are self-dual, i.e. $(\glog,0)^*=(\glog,0)$. 
	This allows us to explore the HT-TS duality in many different directions that 
	go beyond the scope of this paper. 
}. In Eq. (\ref{duality_2}) we see that the unique value of $\nu$, where  $\glog^{*}$ and  $\glog$ and hence $s_{HT}$ and $s_{TS}$ can coincide, is $\nu=0$. 
On one hand, in the $\nu\to0$ limit the HT and the TS entropies are obviously identical. 
On the other hand superstatistical systems are generalizations of BG statistics. It is therefore natural 
for the two approaches to coincide for the BG case. 

\subsection{Example: duality and power laws}
For a large class of distribution functions 
the duality relation of the associated g-logarithms
includes the condition
\begin{equation}
	\glog_{}^*(x)=-\glog_{}(1/x)
	\label{dual_log}
\end{equation}
given in \cite{naudts_physA2002}.
More precisely, a large class of families of 
g-logarithms
is closed and the usual logarithm is self-dual\footnote{Self-duality here means $\log^*(x)=\log(x)$.} 
under this map.
By requiring that both conditions, Eq. (\ref{duality_2}) and 
Eq. (\ref{dual_log}), hold,  
i.e. by inserting Eq. (\ref{dual_log}) into 
Eq. (\ref{duality_2}), it is possible to derive explicitly a 
generic form of 
g-logarithms\footnote{Without loss of generality we can set $c=1$ since $c$ trivially rescales the escort parameter $\nu$.} 
\begin{equation}
\glog_{}(x)=\frac{1}{\frac{1}{\frac{2\lambda}{\nu}\gf\left(\frac{\nu}{2\lambda}\log(x)\right)}-\frac{\nu}{2}}
\quad, 
\label{duality_solution}
\end{equation}
with  $\lambda\in(0,1]$ and $\gf(x)$ a monotonically increasing function $\gf:[-\infty,\infty]\to[-1,1]$, 
with $\gf(0)=0$, $\gf'(0)=1$, $\gf(x)=-\gf(-x)$, and $\lim_{x\to\infty}\gf(x)=1$.
Each pair $\gf$ and $\lambda$ defines a family of g-logarithms
parametrized by $\nu$; each such family $\glog_\nu$ has the properties
(i) $\glog^*_\nu=\glog_{-\nu}$ and (ii) $\lim_{\nu\to0}\glog_\nu=\log$.
The first property states that the g-logarithms associated
with the TS and 
HT entropies are dual to each other. The second property states that in the 
$\nu\to 0$ limit all these families reproduce the BG case\footnote{
	Eq. (\ref{dual_log}) as the starting point for producing $*$-closed
	families of g-logarithms with properties (i) and (ii) is not the most general Ansatz.
	Other possibilities will be discussed elsewhere.
}.
This result is quite general: if $s_{HT}$ is associated with a g-logarithm 
$\glog_{\nu'}$ of the form Eq. (\ref{duality_solution}) for some value $\nu'$ 
then $s_{TS}$ is associated with $\glog_{-\nu'}$, with $\nu'$ fixing the value
of $\nu$ in the escort probability. 

The specific choice of $\gf(x)=\tanh(x)$ and  $\lambda=1$  yields the family of $q$-logarithms
\begin{equation}
	\glog_{}(x) = \frac{1}{\frac{1}{\frac{2}{\nu}\tanh\left(\frac{\nu}{2}\log(x)\right)}-\frac{\nu}{2}} = \frac{x^\nu-1}{\nu} = \log_q(x) \quad, 
	\label{the_q_log}
\end{equation}
where $\log_q(x)=(x^{1-q}-1)/(1-q)$ and $\nu=1-q$.
We see in a concrete example how $\nu$ is uniquely determined 
by the parametrization (here $q$) of the family of given distribution functions.
It is remarkable how requiring 
both,
Eq. (\ref{duality_2}) and Eq. (\ref{dual_log}),
automatically reproduces the condition necessary to recover Tsallis entropy \cite{tsallis88}
within the TS-approach\footnote{
	If a $q$-exponential is observed in the superstatistical system for some particular value of $q$, 
	then the entropy reconstruction in the TS-approach
	produces different entropies $s_{TS}$ depending on the choice of $\nu$ in the TS-MEP. 
	Each of these entropies
	can be used to recover the power-law ($q$-exponential distribution) one has used to construct the
	entropy. 
	Yet only for the particular choice $\nu=1-q$ does one get $\sum_i s_{TS}(p_i)$ as the Tsallis entropy 
	$S_q[{\bf p}]$ for the particular case \cite{TsallisSouza}.
}. 
Moreover, the duality $\glog^*_\nu=\glog_{-\nu}$ written in terms
of $q$-logarithms is just the well-known duality $\log_q^*=\log_{2-q}$. 
The associated HT and TS entropies are
\begin{equation}
\begin{array}{lcl}
S_{TS}[p]&=&-\sum_i\int_0^{p_i} dx\,\log_{2-q}\left(\frac{x}{\rr}\right)\\
&&\\
&=&\frac{1-\sum_i p_i^q}{q-1}\quad,\\
&&\\
S_{HT}[p]&=&-\sum_i\int_0^{p_i} dx\,\log_{q}\left(\frac{x}{\rr}\right)\\
&&\\
&=&-\frac{1}{q(2-q)}\sum_i p_i\log_q(p_i)-\frac{1-q}{q(2-q)}\quad,
\end{array}
\end{equation}
with $\rr=q^{1/(1-q)}$ defined as above by $\left.s_{TS}'(x)\right|_{x=\rr}=0$. 
Clearly $S_{TS}$ gives the Tsallis entropy \cite{tsallisbook} while $S_{HT}$ corresponds to the 
entropy for power-laws discussed in \cite{hanel07,thurner08}.

\section{State changes -- the superstatistical transformation group \label{supergroup}}

As we have seen before, Eq. (\ref{dist}) determines how the superstatistical Boltzmann factor $B$ and the kernel $f$ transform under
state changes $\sigma\to\sigma'$. These transformations are
\begin{equation}
	Z^{(\sigma')}=\frac{1}{z}Z^{(\sigma)} \;\; {\rm and}\;\; B^{(\sigma')}(\ep)=\frac{1}{z}B^{(\sigma)}(a+b\ep) \quad,  
\label{Btrans}
\end{equation}
where $z$ is a ratio of two normalization constants for $\sigma$ and $\sigma'$ respectively. 
For the transformation law of the superstatistical distribution function we find  
\begin{equation}
\begin{array}{lcl}
	\frac{1}{z}B^{(\sigma)}(a+b\ep)&=&\frac{1}{z}\int_0^\infty d\beta f^{(\sigma)}(\beta)e^{-\beta(a+b\ep)}\\
	&=&
	\int_0^\infty d\beta' \left(\frac{1}{z}\frac{1}{b}f^{(\sigma)}(\frac{\beta'}{b})e^{-\beta'\frac{a}{b}}\right)e^{-\beta'\ep}\\
	&=&
	\int_0^\infty d\beta' f^{(\sigma')}(\beta')e^{-\beta'\ep}\\
	&=&
	B^{(\sigma')}(\ep)\,,
\end{array}
\label{affine_1}
\end{equation}
where we substituted $\beta'=b\beta$ in the second line  and used Eq. (\ref{Btrans}) in the third line,  
together with Eq. (\ref{superstat}). By comparison, $f$ transforms according to
\begin{equation}
	f^{(\sigma')}(\beta')=\frac{1}{z}\frac{1}{b}f^{(\sigma)} \left(\frac{\beta'}{b}\right)e^{-\beta'\frac{a}{b}} \quad .
\label{fsig_trans}
\end{equation}
The value of $z$ is fixed by the normalization condition for the kernel, $\int_0^\infty d\beta f^{(\sigma')}(\beta)=1$.

For describing the transformation group we introduce the following notation: state $\sigma$ is characterized by $a=0$ and $b=1$.
For state $\sigma'$, $a$ and $b$ take other values. We now indicate the parameters $(a,b)$ in the kernel through 
\begin{equation}
	f^{(\sigma)}\equiv f^{(0,1)}\quad,\quad f^{(\sigma')}\equiv f^{(a,b)} \quad.
\end{equation}
Equation (\ref{fsig_trans}) can now be written as 
\begin{equation}
	f^{(a,b)}(\beta')=\frac{1}{z}\frac{1}{b}f^{(0,1)} \left(\frac{\beta'}{b}\right)e^{-\beta'\frac{a}{b}}\quad.
\label{fsig_trans_ab}
\end{equation}
From Eq. (\ref{affine_1}) we see that under the state change $\sigma\to\sigma'$ the energy 
$\ep$ undergoes the affine transformation $\ep\to a+b\ep$.
We define three operators $\Phi_1$, $\Phi_2$, and $\Pi$ representing translation of $\ep$, dilatation of $\ep$, and  normalization of $f$, respectively
\begin{equation}
\begin{array}{lcl}
	\Phi_1(a)f(\beta)&=&e^{-\beta a}f(\beta)\\
	\Phi_2(b)f(\beta)&=&\frac{1}{b}f(\frac{\beta}{b})\\
	\Pi f(\beta)&=&\frac{f(\beta)}{\int_0^\infty d\beta'f(\beta')}\quad.
\end{array}
\label{affine_2}
\end{equation}
$\Phi_1$ and $\Phi_2$ form groups 
\begin{equation}
\begin{array}{rcl}
	\Phi_1(a)\Phi_1(a')&=&\Phi_1\left(a+a'\right)\\
	\Phi_2(b)\Phi_2(b')&=&\Phi_2\left(bb'\right) \quad,
\end{array}
\label{affine_3}
\end{equation}
with the identity elements $\Phi_1(0)$, and $\Phi_2(1)$, and the  inverse elements
\begin{equation}
	\Phi_1^{-1}(a)=\Phi_1\left(-a\right)\quad,\quad\Phi_2^{-1}(b)=\Phi_2\left(\frac{1}{b}\right) \quad.
\label{affine_4}
\end{equation}
$\Phi_1(a)$ and $\Phi_2(b)$ do not commute: 
\begin{equation}
	\Phi_2(b)\Phi_1(a)=\Phi_1\left(\frac{a}{b}\right)\Phi_2(b)\quad ,
\label{affine_5}
\end{equation}
while the projection operator $\Pi$ commutes with $\Phi_2$.
We finally  define the group
\begin{equation}
	G(a,b)=\Pi\,\, \Phi_2(b) \,\, \Phi_1(a) \quad ,
\label{affine_6}
\end{equation}
with the following group-composition rule and inverse element
\begin{equation}
\begin{array}{l}
	G(a,b)G(a',b')=G(a'+ab',bb')\\
	G^{-1}(a,b)=G\left(-\frac{a}{b},\frac{1}{b}\right)\quad.
\end{array}
\label{affine_7}
\end{equation}
Equation  (\ref{fsig_trans_ab}) can now be written as
\begin{equation}
	f^{(a,b)}(\beta)=G(a,b) \,\, f^{(0,1)}(\beta)\quad, 
\label{f_trans_G}
\end{equation}
or more generally,
\begin{equation}
	f^{(a',b')}(\beta)=G\left(\frac{a'-a}{b},\frac{b'}{b}\right) \,\, f^{(a,b)}(\beta)\quad.
\label{f_trans_GG}
\end{equation}
The generators of $\Phi_1$ and $\Phi_2$ are given 
by $g_1=-\beta$ and  $g_2=-(1+\beta \frac{d}{d\beta})$. 
We have a representation of the Euclidean group in one dimension.

These transformations apply to infinitely many different $f$'s. For example, if $f^{(0,1)}$ leads to a 
$q$-exponential distribution with a specific value of $q$, under these transformations all possible $f^{(a,b)}$
will give $q$-exponentials with the same value of $q$. 
The functional form of distribution functions is preserved under the transformation group.
More generally, for every type of superstatistical distribution function, we can identify 
a representative $f^{(0,1)}$ and call it the `progenitor'  function. 
From this a sequence 
$\Xi(f)$ 
of $f^{(a,b)}$ can be generated by applying the transformation group. For every type of function  
$f$ there exists a unique sequence to which it belongs. These sequences define equivalence classes for the superstatistical kernels. 
Suppose experiments on a superstatistical system produce a set of functions $\Omega=\{f_{\sigma_i}\}_{i=1}^N$
for $N$ different states $\sigma_i$, arbitrarily chosen. If $\Omega\subset \Xi(f)$ for some fixed sequence $\Xi(f)$, then state changes of the system transform consistently under the transformation group Eq. (\ref{affine_6})
and an entropy $S=\sum_i s(p_i)$ exists such that $s$
is invariant under state changes of the superstatistical system.

\section{Discussion}

By studying the possibilities of reconstructing entropies from superstatistical distribution functions we found that there exist only two distinct frameworks,
the HT approach (using only normal constraints) and the TS approach (employing escort constraints).
A duality between the two approaches is shown to exist.
This indicates that superstatistical systems can be explored in two  ways which are {\em dual} to each other.
Under certain circumstances the duality allows us to determine the possibilities of functional forms of generalized logarithms and thus 
of entropic forms, as explicitly shown for systems with characteristic power laws.
For systems governed by a particular distribution function it has to be decided empirically which of the two frameworks applies; these systems may differ considerably in terms of their thermodynamic properties. 
We studied how the 
superstatistical kernel transforms under macroscopic state changes within the two approaches. 
We presented the respective transformation laws and identified the corresponding transformation 
group to be the Euclidean group in one dimension. This indicates the existence of a remarkably simple mathematical structure 
behind thermodynamical state changes of superstatistical systems 
governed by an invariant entropic form.

\begin{acknowledgments}
R.H. and S.T. thank the SFI for hospitality. M. G.-M. is glad to acknowledge the generous support of Mr. Jerry Murdock and the Bryan J. and June B. Zwan Foundation. In addition, he is greatly indebted to Professor E. G. D. Cohen for valuable conversations.
\end{acknowledgments}

\end{article}
\end{document}